\documentclass[11pt]{article}
\usepackage[dvips]{graphics}
\def\ra{\rightarrow}
\def\iy{\infty}

\def\be{\begin{equation}}
\def\ee{\end{equation}}
\def\ov{\over}
\def\ve{\varepsilon}
\begin{document}

\title{Airy Kernel and Painlev\'e II}

\author{Craig A.~Tracy\\
Department of Mathematics\\ and \\
Institute of Theoretical Dynamics\\
University of California\\
 Davis, CA 95616, USA\\
email address: tracy@itd.ucdavis.edu
\vspace{2ex}
\\
Harold Widom\\
Department of Mathematics\\
University of California\\
Santa Cruz, CA 95064, USA\\
email address: widom@math.ucsc.edu
}
\date{}
\maketitle
\abstract{We prove that the distribution function
of the largest eigenvalue in the Gaussian Unitary
Ensemble (GUE) in the edge scaling limit is expressible
in terms of Painlev\'e II.   Our goal is to
concentrate on this important example of the connection
between random matrix theory and integrable systems,
and in so doing to introduce
the newcomer to the subject as a whole.
We also give sketches of the results for the limiting
distribution of the largest eigenvalue
 in the  Gaussian Orthogonal Ensemble (GOE) and
 the Gaussian Symplectic Ensemble (GSE).
This work we did  some
years ago in a more general setting.  These notes, therefore, are
\textit{not} meant for experts in the field.
}

\section{Introduction}
For the finite $N$ random matrix models GOE, GUE and GSE,
the probability that no eigenvalues lie in a set $J$, $E_{\beta}(0;J)$,
equals a Fredholm determinant of an integral operator
with scalar kernel for GUE ($\beta=2$) and equals the square root
of a Fredholm determinant of an integral operator
with matrix kernel for GOE ($\beta=1$) and GSE ($\beta=4$).
Gaudin~\cite{gaudin} (using Mehta's newly invented method
of orthogonal polynomials~\cite{mehtaPoly})
 first discoverd this connection between random matrix
theory and Fredholm determinants, for GUE in
the bulk scaling limit,  in 1961.
In 1970 Dyson~\cite{dyson}
showed that the $n$-point correlations for circular
ensembles are expressible as
quaternion determinants.  The next
year Mehta~\cite{mehta1}  extended Dyson's formalism to include
the GOE and the GSE.  From these representations of the
$n$-point correlations, one  deduces that  $E_\beta(0;J)^2$
($\beta=1,4$) are  Fredholm determinants.  Later
several authors  generalized this construction to other invariant
ensembles. See~\cite{correlations,w1} for simplified
proofs of these facts.  
\section{Distribution of Largest Eigenvalue, $\beta=2$}
\subsection{Fredholm Determinant Representation}
In finite $N$ GUE (see, e.g., \cite{mehta,correlations})
\[ E_{N,2}(0;J)=\det\left(I-K_{N,2}\right)\]
where $K_{N,2}$ is the integral operator
\[ \left(K_{N,2}f\right)(x)=\int_J K_{N,2}(x,y) f(y)\, dy, \]
\begin{eqnarray}
 K_{N,2}(x,y) &=& {\varphi(x) \psi(y) - \psi(x) \varphi(y)\ov
x-y},\label{kernel} \\
 \varphi(x)&= &\left(N\ov 2\right)^{1/4}\varphi_N(x),\label{hermite} \\
 \psi(x)&=&\left(N\ov 2\right)^{1/4}\varphi_{N-1}(x)\label{hermite2}
 \end{eqnarray}
 and the $\varphi_k(x)$ are the harmonic oscillator wave functions
 \[ \varphi_k(x) = {1\over \sqrt{2^k k! \sqrt{\pi}}}\, e^{-x^2/2} H_k(x). \]
 (The $H_k(x)$ are the Hermite polynomials and the sequence
$\{\varphi_k\}_{k\ge 0}$
 forms a complete orthonormal sequence in $L^2(\textbf{R})$.)

 For $J=(t,\infty)$, the probability that no
 eigenvalues lies in $J$ is the same as the probability that all
 eigenvalues lie to the left of $t$;  and in particular, that
 the largest eigenvalue, $\lambda_{\textrm{max}}(N)$, lies to the
 left of $t$:
 \[ F_{N,2}(t):=\textrm{Prob}\left(\lambda_{\textrm{max}}(N)< t\right)=
 E_2\left(0;(t,\infty)\right). \]
 (Since the distribution functions all have continuous densities,
 we need not worry about the
 difference between $<$ and $\le$.)
 \subsection{Edge Scaling Limit}
 \textit{Fixing} $t$, $\textrm{Prob}\left(\lambda_{\textrm{max}}(N)< t\right)
 \rightarrow 0$ as
 $N\rightarrow\infty$ since the
 largest eigenvalue increases with $N$; and hence, the probability that it
remains less than
 any fixed number tends to zero.  To obtain, therefore, a nontrivial limiting
distribution we must
 introduce a normalized random variable.  Such situations are common in
 probability theory, e.g.\  the central limit theorem.

The density of eigenvalues at $x$, $\rho(x)$,  equals $K_{N,2}(x,x)$.
Replacing $K_{N,2}(x,y)$ by
\[ {1\ov \rho(z)} K_{N,2}\left(z+{x\ov\rho(z)},z+{y\ov\rho(z)}\right)\]
has the effect of making the point $z$ in the spectrum the new origin and
rescaling
so that the eigenvalue density at this point becomes equal to $1$.  The ``edge
of the spectrum''
corresponds to $z\sim\pm\sqrt{2N}$.  (The constant $\sqrt{2}$ results from  the
choice
of the standard deviation of the matrix elements in GUE---here we choose  a
standard
deviation equal to $1/\sqrt{2}$ which is the same as
in Mehta~\cite{mehta}.)
The \textit{edge scaling limit}~\cite{bowick,forrester} for the kernel  is then
\be
 \lim_{N\rightarrow\infty}{1\ov 2^{1/2} N^{1/6}}\, K_{N,2}\left(\sqrt{2N}+{x\ov
2^{1/2} N^{1/6}},
\sqrt{2N}+{y\ov 2^{1/2} N^{1/6}}\right)=K_{\textrm{Airy}}(x,y)
\label{airyKernel}\ee
where $K_{\textrm{Airy}}$ has the form (\ref{kernel}) with
\[ \varphi(x)=\textrm{Ai}(x), \ \  \psi(x)= \textrm{Ai}^\prime(x),\]
and $\textrm{Ai}(x)$  the Airy function.  Given $t$ and $N$, we  define $s$ by
\[ t= \sqrt{2N}+ {s\ov 2^{1/2} N^{1/6}}. \]
Using  (\ref{airyKernel}) and the Fredholm
expansion of $\det(I-K_{N,2})$, it follows  as $N\rightarrow\infty$ and
$t\rightarrow\infty$ such that $s$  is fixed that (the  \textit{edge scaling
limit})
\[ \lim F_{N,2}(t)=F_2(s)=\det\left(I-K_{\textrm{Airy}}\right) \]
where $K_{\textrm{Airy}}$ acts on $L^2\left((s,\infty)\right)$.

\noindent\textbf{Remarks:}  Inspecting (\ref{hermite}) and
(\ref{hermite2}) we see that one needs asymptotic formulas for the Hermite
polynomials
$H_k(x)$ for large
values of both $x$ and $k$.  Fortunately, such  formulas are known and were
derived by
Plancherel and Rotach in 1929.   (They can be found in  the Bateman
Manuscripts.)
Using these asymptotic formulas it is then straighforward to derive
(\ref{airyKernel}).
Recent progress has been made in
deriving analogous Plancherel-Rotach type formulas for orthogonal polynomials
whose weight function
is of the form $e^{-V}$~\cite{bleher,deiftPoly}.  The proofs are considerably
more difficult
than in the classical Hermite case.  These formulas then allow one to prove
that the
Airy kernel arises in the edge scaling limit
 under much more general circumstances.

 \subsection{Painlev\'e II Representation for $F_2$}
For notational convenience we denote $K_{\textrm{Airy}}$ by $K$ in this
subsection.
It will also be convenient to think of our operator $K$ as acting, not on
$(s,\infty)$, but
on \textbf{R} and to have kernel
\[ K(x,y) \chi(y) \]
where $\chi$ is the characteristic function of $J$.  Since the integral
operator $K$ is trace-class and depends smoothly on
the parameter $s$, we have  the well known formula
\be {d\ov ds}\log\det\left(I-K\right)=-\textrm{tr}\left(\left(I-K\right)^{-1}
 {\partial K\ov \partial s}\right). \label{dLog}\ee
By calculus
\be {\partial K\ov \partial s}\doteq -K(x,s)\delta(y-s). \label{Kderiv}\ee
(If $L$ is an operator with kernel $L(x,y)$ we denote this by $L\doteq
L(x,y)$.)
Substituting this into the above expression gives
\[ {d\ov ds} \log\det\left(I-K\right)= - R(s,s)\]
where $R(x,y)$ is the resolvent kernel of $K$, i.e.\  $R=(I-K)^{-1}K\doteq
R(x,y)$. The resolvent
kernel $R(x,y)$ is smooth in $x$ but discontinuous in $y$ at $y=s$.  The
quantity
$R(s,s)$ is interpreted to mean
\[ \lim_{{y\rightarrow s \atop y\in J}}R(s,y).\]

\subsubsection{Representation for $R(x,y)$}
If $M$ denotes the multiplication operator, $(Mf)(x)=x f(x)$,  then
\[ \left[M,K\right]\doteq x K(x,y)- K(x,y) y = (x-y) K(x,y) =
 \varphi(x) \psi(y) - \psi(x) \varphi(y). \]
As an operator equation this is
\[ \left[M,K\right]=\varphi\otimes \psi - \psi\otimes \varphi. \]
(We define  $a\otimes b\doteq a(x) b(y)$ and let $\left[\cdot,\cdot\right]$
denote the commutator.)
Thus
\begin{eqnarray}
\left[M,\left(I-K\right)^{-1}\right]&=&\left(I-K\right)^{-1} \left[M,K\right]
 \left(I-K\right)^{-1}
 \nonumber \\
 	&=&\left(I-K\right)^{-1}\left(\varphi\otimes \psi - \psi\otimes
\varphi\right)
	\left(I-K\right)^{-1}\nonumber\\
	&=& Q\otimes P - P\otimes Q \label{comm1}
	\end{eqnarray}
where we have introduced
\be Q(x;s)=Q(x)= \left(I-K\right)^{-1} \varphi \ \ \ \textrm{and} \ \ \
P(x;s)=P(x)=
 \left(I-K\right)^{-1}
\psi. \label{QP}\ee
On the other hand since $(I-K)^{-1}\doteq \rho(x,y)=\delta(x-y)+R(x,y)$,
\be \left[M,\left(I-K\right)^{-1}\right]\doteq (x-y)\rho(x,y)=(x-y) R(x,y).
\label{comm2}\ee
Comparing (\ref{comm1}) and (\ref{comm2}) we see that
\be
R(x,y) = {Q(x) P(y) - P(x) Q(y) \ov x- y}, \ \ x,y\in J. \label{R} \ee

Taking $y\rightarrow x$ gives
 \be R(x,x)= Q^\prime(x) P(x) - P^\prime(x) Q(x). \label{Rdiag} \ee
Introducing
\be q(s)=Q(s;s) \ \ \ \textrm{and} \ \ \ p(s) = P(s;s), \label{qp} \ee
we have
\be R(s,s) = Q^\prime(s;s) p(s) - P^\prime(s;s) q(s),\ \  s<x,y<\infty.
\label{RDiag}\ee

The reader may have noted that the expression (\ref{R}) for $R(x,y)$  depends
only upon the kernel $K(x,y)$ having the form (\ref{kernel}).  As far as the
authors
are aware, the generality of this representation for $R(x,y)$ was first
stressed
by Its, et al.~\cite{its1}, though it appears in the context of the sine kernel
in the  work of Jimbo, et al.~\cite{jmms}.

\subsubsection{Formulas for $Q^\prime(x)$ and $P^\prime(x)$}
As we just saw, we need expressions for $Q^\prime(x)$ and $P^\prime(x)$. If
$D$ denotes the differentiation operator, $d/dx$,  then
\begin{eqnarray}
Q^\prime(x;s)&=& D \left(I-K\right)^{-1} \varphi \nonumber\\
&=& \left(I-K\right)^{-1} D\varphi +
\left[D,\left(I-K\right)^{-1}\right]\varphi\nonumber\\
&=& \left(I-K\right)^{-1} \psi +
\left[D,\left(I-K\right)^{-1}\right]\varphi\nonumber\\
&=& P(x) + \left[D,\left(I-K\right)^{-1}\right]\varphi. \label{Qderiv1}
\end{eqnarray}
We need the commutator
\[ \left[D,\left(I-K\right)^{-1}\right]=\left(I-K\right)^{-1} \left[D,K\right]
\left(I-K\right)^{-1}. \]
Integration by parts shows
\[ \left[D,K\right] \doteq \left( {\partial K\ov \partial x} + {\partial K\ov
\partial y}\right)
+ K(x,s) \delta(y-s). \]
(The $\delta$ function comes from differentiating the
characteristic function $\chi$.)
Using the specific form for $\varphi$ and $\psi$  ($\varphi^\prime=\psi$,
$\psi^\prime=x\varphi$)
we compute:
\[ \left( {\partial K\ov \partial x} + {\partial K\ov \partial y}\right) =
\varphi(x) \varphi(y). \]
Thus
\be \left[D,\left(I-K\right)^{-1}\right]\doteq - Q(x) Q(y) + R(x,s) \rho(s,y).
\label{DComm}\ee
(Recall $(I-K)^{-1}\doteq \rho(x,y)$.)  We now use this in (\ref{Qderiv1})
\begin{eqnarray*}
 Q^\prime(x;s)&=&P(x) - Q(x) \left(Q,\varphi\right) + R(x,s) q(s) \\
 &=& P(x) - Q(x) u(s) + R(x,s) q(s)
 \end{eqnarray*}
 where the inner product $\left(Q,\varphi\right)$ is denoted by $u(s)$.
 Evaluating  at $x=s$  gives
 \be Q^\prime(s;s) = p(s) - q(s) u(s) +R(s,s) q(s). \label{Qderiv2} \ee

 We now apply the same procedure to compute $P^\prime$ encountering the one
 new feature that since $\psi^\prime(x)=x\varphi(x)$ we need
 to introduce an additional commutator term:
 \begin{eqnarray*}
 P^\prime(x;s)&=& D \left(I-K\right)^{-1} \psi \\
 &=& \left(I-K\right)^{-1} D\psi + \left[D,\left(I-K\right)^{-1}\right]\psi\\
 &=& M \left(I-K\right)^{-1} \varphi +
\left[\left(I-K\right)^{-1},M\right]\varphi+
 \left[D,\left(I-K\right)^{-1}\right]\psi\\
 &=& x Q(x) +\left(P\otimes Q-Q\otimes P\right)\varphi +(-Q\otimes Q)\psi +
R(x,s) p(s)\\
 &=& x Q(x) + P(x)\left(Q,\varphi\right) -  Q(x) \left(P,\varphi\right)
 - Q(x) \left(Q,\psi\right)+R(x,s)p(s)\\
 &=& x Q(x) - 2 Q(x) v(s) + P(x) u(s) + R(x,s) p(s).
 \end{eqnarray*}
 Here $v=\left(P,\varphi\right)=\left(\psi,Q\right)$.
 Evaluating  at $x=s$ gives
 \[ P^{\prime}(s;s) = s q(s) + 2 q(s) v(s) +p(s) u(s) +R(s,s) p(s).
\label{Pderiv}\]
 Using this and the expression for $Q^\prime(s;s)$ in (\ref{RDiag})
 gives
 \be R(s,s)= p^2-s q^2 + 2 q^2 v - 2 p q u. \label{RDiag2} \ee

 \subsubsection{First order equations for $q$, $p$, $u$ and $v$}
 By the chain rule
 \be {dq\ov ds} = \left( {\partial\ov \partial x}+{\partial\ov \partial
s}\right)
  Q(x;s)\left\vert_{x=s}. \right.
 \label{qDeriv}\ee
 We have already computed the partial of $Q(x;s)$ with respect to $x$.
 The partial with respect to $s$ is
 \begin{eqnarray*}
  {\partial Q(x;s)\ov \partial s}&=& \left(I-K\right)^{-1} {\partial K\ov
\partial s}
   \left(I-K\right)^{-1} \varphi\\
  &=& - R(x,s) q(s)
  \end{eqnarray*}
  where we used (\ref{Kderiv}).  Adding the two partial derivatives  and
evaluating at $x=s$
  gives
  \be {dq\ov ds} = p - q u. \label{qEqn}\ee
  A similar calculation gives
  \be {dp\ov ds}= s q - 2 q v + p u. \label{pEqn} \ee
  We derive first order differential equations for  $u$ and $v$  by
differentiating
  the inner products:
  \begin{eqnarray*}
  u(s) &=& \int_s^\infty \varphi(x) Q(x;s)\, dx, \\
  {du\ov ds}&=& -\varphi(s) q(s) + \int_s^\infty \varphi(x) {\partial Q(x;s)\ov
\partial s}\, dx \\
  &=& -\left(\varphi(s)+\int_s^\infty R(s,x) \varphi(x)\,dx\right) q(s)\\
  &=& -\left(I-K\right)^{-1} \varphi(s) \, q(s)\\
  &=& - q^2.
  \end{eqnarray*}
  Similarly,
  \[ {dv\ov ds} = - p q. \]

  \subsubsection{Painlev\'e II and $F_2$}
  From the first order differential equations for $q$, $u$ and $v$ it follows
immediately
  that the derivative of
  $ u^2-2v-q^2 $ is zero.  Examining the behavior near $s=\infty$ to check that
  the constant of integration is zero then gives
  \[ u^2-2v=q^2, \]
  a \textit{first integral}.
 We now differentiate (\ref{qEqn}) with respect to $s$, use the first order
differential
  equations for $p$ and $u$, and then
  the  first integral to deduce that $q$ satisfies the \textit{Painlev\'e II}
equation
  \be q^{\prime\prime}=s q + 2 q^3. \label{P2}\ee
  Checking the asymptotics of
  the Fredholm determinant $\det(I-K)$ for large $s$ shows we want the solution
to
  the Painlev\'e II equation with boundary condition
  \be q(s)\sim \textrm{Ai}(s) \ \ \ \textrm{as} \ \ \ s\rightarrow\infty.
\label{bc} \ee
That a solution $q$ exists and is unique follows from the representation
of the Fredholm determinant in terms of it.  Independent proofs of this,
as well as the asymptotics as $s\ra\-\iy$ were given by \cite{hastings,
clarkson,  deiftP2}.

  Since the kernel of $[D,(I-K)^{-1}]$ is $(\partial/\partial
x+\partial/\partial y)R(x,y)$,
  (\ref{DComm}) says
  \[ \left({\partial\ov \partial x}+{\partial\ov\partial
y}\right)R(x,y)=-Q(x)Q(y)+R(x,s)\rho(s,y). \]
  In computing $\partial Q(x;s)/\partial s$ we showed that
  \[ {\partial\ov \partial s} \left(I-K\right)^{-1}\doteq {\partial\ov\partial
s}R(x,y)
   = -R(x,s)\rho(s,y). \]
  Adding these two expressions,
  \[ \left({\partial\ov\partial x}+{\partial\ov\partial y}+
  {\partial\ov\partial s}\right)R(x,y)=-Q(x) Q(y), \]
  and then evaluating at $x=y=s$ gives
  \be {d\ov ds}R(s,s)=-q^2. \label{Rderiv} \ee
  Integration  (and recalling (\ref{dLog})) gives,
   \[ {d\ov ds}\log\det\left(I-K\right)=-\int_s^\infty q^2(x) \, dx; \]
  and hence,
  \[ \log\det\left(I-K\right)=-\int_s^\infty\left(\int_y^\infty
q^2(x)\,dx\right)\, dy
  =-\int_s^\infty (x-s) q^2(x)\, dx. \]

  To summarize,  we have shown that the distribution function $F_2$ has the
Painlev\'e
  representation
  \be F_2(s) = \exp\left(-\int_s^\infty (x-s) q^2(x)\, dx\right) \label{F2} \ee
  where $q$  satifies the Painlev\'e II equation (\ref{P2}) subject to
  the boundary condition (\ref{bc}).

  Alternatively, it is easy to check that (\ref{Rderiv}) integrates to
  \be R(s,s)=(q^\prime)^2-s q^2 - q^4. \label{Rq} \ee
  This  and together with (\ref{P2}) shows that $R(s,s)$ itself satisfies
  \[ (R^{\prime\prime})^2+4 R^\prime\left((R^\prime)^2-s R^\prime +R\right)=0.
\]
  This  last differential equation is the Jimbo-Miwa-Okamoto~\cite{jm,okamoto}
$\sigma$ form for
  Painlev\'e II.  (Observe that (\ref{Rq}) and (\ref{RDiag2})
  give another identity between $q$, $p$, $u$ and $v$.)

  \subsection{Remarks}
  \begin{enumerate}
  \item  In~\cite{airy} the general case $J=\bigcup(a_{2j-1},a_{2j})$ is first
considered
  and then specialized to  $J=(s,\infty)$.  Here we have simply given the proof
  of \cite{airy} for this special case from the very beginning.

  \item The asymptotics of $F_2(s)$ as $s\rightarrow -\infty$ are important and
  they require the
  solution of a connection problem for Painlev\'e II.  This connection problem
is solved in~\cite{hastings, clarkson, deiftP2} and applied in~\cite{airy} to
$F_2$ with the result that
  \[ F_2(s)\sim {\tau_0\ov (-s)^{1/8}}\, e^{s^3/12}, \ \ s\rightarrow -\infty.
\]
  The constant $\tau_0$ is conjectured to equal  $e^{\zeta^\prime(-1)}
2^{1/24}$.

  \item For the distribution of the next-largest, next-next-largest, etc.\
eigenvalues,
  see~\cite{airy}.

  \item  The connection between certain classes of Fredholm determinants and
integrable
  systems of the Painlev\'e type begins with work on the two-dimensional Ising
model~\cite{wmtb,mtw}.
  In this model the spin-spin correlation functions in the scaling limit are
expressed
  in terms of a third Painlev\'e transcendent.  Painlev\'e functions  first
appear
  in random matrix theory in~\cite{jmms} where  the Fredholm
  determinant of the sine kernel, $J=(-t,t)$, is
  given in terms of a fifth Painlev\'e transcendent.

  \item  In~\cite{fredholm} a general theory
  was developed that relates Fredholm determinants of operators
  $K$ with kernel of the form (\ref{kernel}) acting on $J=\bigcup
(a_{2j-1},a_{2j})$ with integrable systems.   This theory assumes that
  the functions $\varphi$ and $\psi$ satisfy a first order linear system of
differential equations with rational coefficients.
For related developments see~\cite{adlerShiota, harnad, palmer, witte}.

  \end{enumerate}

 \section{Orthogonal and Symplectic Ensembles}

 The probabilities, $E_\beta(0;J)$, that a set $J$ consisting
 of a finite union of open intervals contains no eigenvalues for
 the finite $N$ GOE ($\beta=1$) and GSE ($\beta=4$) are  more difficult
 to compute
  than are the probabilities for the finite $N$ GUE ($\beta=2$).   The
essential
 problem comes down to the fact that we must deal with Fredholm determinants
  of operators with \textit{matrix-valued} kernels~\cite{orthogonal}.

  Changing notation slightly, we denote by $S$
  the operator $K_{N,2}$ of the previous section and let $\varphi$
  and $\psi$ be as before; namely,  (\ref{hermite})
  and (\ref{hermite2}), respectively.  Let $\ve$ equal the operator
  with kernel $\ve(x-y)={1\ov 2} \textrm{sgn}(x-y)$ ($sgn$ is the
  sign function) and  $D$   the
  differentiation operator.  It then follows from Mehta's
representation~\cite{mehta1} of
  the $n$-point correlations~\cite{orthogonal,correlations} and  for $N$ even
(this
  case is slightly simpler) that
  \[ E_{N,1}(0;J)^2 = \det\left(I-K_{N,1}\right)\]
  where
  \[ K_{N,1}=\chi\left(
  \begin{array}{cc}
  S+\psi\otimes \ve\varphi & SD-\psi\otimes \varphi \\
  \ve S - \ve + \ve\psi\otimes \ve\varphi & S+\ve\varphi\otimes\psi
  \end{array}
  \right)\chi \]
  and $\chi$ is the operator of multiplication by $\chi_J(x)$, the
characteristic
  function of $J$.  There is a similar representation
  for $E_{N,4}(0;J)$~\cite{orthogonal,correlations}.
  This form for $K_{N,1}$ indicates our change of view; namely, we think of
$K_{N,1}$
  as a $2\times 2$ matrix with operator entries.  We then manipulate these
determinants
  of these operators
  to put  them in the form
  \[ E_{N,\beta}(0;J)^2 = E_{N,2}(0;J) \, \det\left(I-\sum_{k=1}^n
\alpha_k\otimes \beta_k\right) \]
  for $\beta=1,4$.  (Of course, the $\alpha_k$'s and $\beta_k$'s depend upon
the ensemble.)
 The last determinant is evaluated using the general formula~\cite{gohberg}
 \[ \det\left(I-\sum_{k=1}^n \alpha_k\otimes \beta_k\right)=
 \det\left(\delta_{j,k}-(\alpha_j,\beta_k)\right)_{j,k=1,\ldots,n} \]
 where $(\alpha_j,\beta_k)$ denotes the inner product.

 We then derive differential equations for the inner products introduced by the
last
 determinant.  For $J=(t,\infty)$, $E_\beta(0;J)$  is again the
 distribution function of the largest eigenvalue, $F_{N,\beta}(t)$, in
 either the GOE ($\beta=1$) or GSE ($\beta=4$).
 Again we take the edge scaling limit:
 \[ F_\beta(s):=\lim_{N\rightarrow\infty} F_{N,\beta}\left(2\sigma\sqrt{N}+
 {\sigma s\ov N^{1/6}}\right).\]
 Here $\sigma$ is the standard deviation of the Gaussian distribution on the
 off-diagonal matrix elements.  Our choice corresponds to $\sigma=1/\sqrt{2}$,
but as
 the notation suggests, $F_\beta(s)$ is independent of $\sigma$.  Our final
results are
 \begin{eqnarray}
 F_1(s)^2&=&F_2(s) \exp\left(-\int_s^\infty q(x)\, dx\right), \label{F1}\\
 F_4(s/\sqrt{s})^2&=& F_2(s) \cosh^2\left({1\ov 2}\int_s^\infty q(x)\,
dx\right)\label{F4}
 \end{eqnarray}
 and $F_2$ and $q$ are as before.

\subsection{Remarks}
\begin{enumerate}
\item Strictly speaking these distributions are only applicable in the
limit that the size of the matrices, $N$, tends to infinity.  In practice
they are good approximations once $N\ge 200$.  Some comparisons
with finite $N$ simulations are in~\cite{montreal}.

\item Table 1 displays some statistics of $F_\beta$ and
 Figure 1 graphs the densities $f_\beta(s)=dF_\beta/ds$.

\item  As has been established for the edge scaling limit of unitary
matrix ensembles~\cite{bleher, deiftPoly, soshnikov}, we expect the distributions
$F_1$ and $F_4$ to be the generic distributions at the edge of the spectrum
(no ``fine tuning'' of the potential) for a large class of orthogonal
and symplectic matrix ensembles. This universality has been proved for the
orthogonal and symplectic ensembles in the case of a quartic potential~\cite{stojanovic}.
In~\cite{johnstone} the distribution of the appropriately centered and normalized
largest eigenvalue of a $p$ variate Wishart distribution
on $n$ degrees of freedom is shown to converge, as $n\ra\iy$, $p\ra\iy$,
such that $n/p$ is fixed, to $F_1$.

\item The past few years has seen an explosion of activity connecting
random matrix theory and combinatorial problems related to the
Robinson-Schensted-Knuth algorithm.  At the level of
limit laws, this  began with the work of
Baik, Deift and Johansson~\cite{bdj} (for a review of this
work, see~\cite{deift}) who discovered
 that  $F_2$ is the  limiting distribution of the
 (appropriately centered and normalized) length of the longest increasing subsequence
 of a random permutation.  For related developments
see~\cite{adlervanM, aldousDiaconis,  baikRains1, baikRains2,
 baikRains3, borodin, gtw1, gtw2, itw1, itw2,
johansson1, johansson2, kuperberg,  okounkov,praehofer1, praehofer2, 
signed, words, vanM}.
\end{enumerate}

\noindent{\textbf{Acknowledgements}

The first author thanks Oleg Zaboronski and Alan Newell
at the Mathematics Institute, University of Warwick,
and John Harnad and Alexander Its at the CRM Workshop
\textit{Isomondromic Deformations and Applications} for their
kind hospitality. 
 This work was supported in part by the National Science Foundation
 through grants DMS--9802122 (first author) and DMS--9732687 (second
 author).

\begin{table}
\begin{center}
\caption{The mean ($\mu_\beta$),  standard deviation ($\sigma_\beta$),
skewness ($S_\beta$) and  kurtosis ($K_\beta$) of $F_\beta$.}
\vspace{4ex}
\begin{tabular}{|l|cccc|}\hline
$\beta$ & $\mu_\beta$ & $\sigma_\beta$ & $S_\beta$ & $K_\beta$ \\  \hline
1 & -1.20653 & 1.2680 & 0.293 & 0.165 \\
2 & -1.77109 & 0.9018 & 0.224 & 0.093 \\
4 & -2.30688 & 0.7195 & 0.166 & 0.050 \\ \hline
\end{tabular}
\end{center}
\end{table}

\begin{figure}
\vspace{0in}
\begin{center}
\caption{The probability density $f_\beta$
 of the largest eigenvalue, $\beta=1,2,4$.}
\vspace{2ex}
\resizebox{9cm}{9cm}{\includegraphics{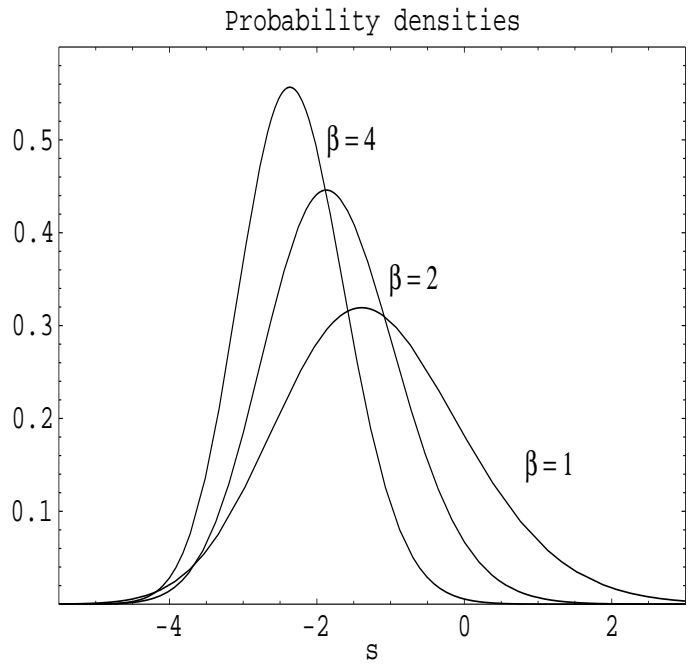}}
\end{center}
\end{figure}

\end{document}